\documentclass[a4paper,11pt]{article}
\usepackage{jcappub}
\pdfoutput=1 

\usepackage[utf8]{inputenc}
\DeclareUnicodeCharacter{223C}{\textasciitilde}

\usepackage{psfrag,color,epsfig}
\usepackage{graphicx,graphics}	
\usepackage{upgreek}
\usepackage{amssymb}	
\usepackage{multicol}	
\usepackage{bm}		
\usepackage{pdflscape}	
\usepackage{xspace}
\usepackage[T1]{fontenc} 
\usepackage{subfig}
\usepackage[multiple]{footmisc}
\usepackage{soul}

\definecolor{darkgreen}{rgb}{0, 0.70, 0}
\usepackage{orcidlink}

\title{ Towards a Robust Machine-Learning Pipeline for 21-cm Cosmology Data Analysis I: A Roadmap for Development and Demonstration of Robustness Against PSF Modeling Errors}

\author[\orcidlink{0000-0003-4395-4931}, a]{Madhurima Choudhury}

\author[\orcidlink{0000-0002-3492-0433},a,b]{Jonathan C. Pober}
\affiliation[a]{Center for the Fundamental Physics of the Universe, Brown University \\
184 Hope St \\
Providence, RI 02912, USA}
\affiliation[b]{Department of Physics, Brown University \\
184 Hope St \\
Providence, RI 02912, USA}

\emailAdd{madhurimachoudhury811@gmail.com}
\emailAdd{jonathan\_pober@brown.edu}

\abstract{
The 21-cm signal from the Epoch of Reionization (EoR) is a powerful probe of the evolution of the Universe. However, accurate measurements of the EoR signal from radio interferometric observations are sensitive to efficient foreground removal, mitigating radio-frequency interference and accounting for instrumental systematics. This work represents the first in a series of papers, where we will be introducing a novel ML based pipeline, step-by-step, to directly infer reionization parameters from 21-cm radio-interferometric images.
In this paper, we investigate the impact of the variations in the point spread function (PSF) on parameter estimation by simulating visibilities corresponding to input 21-cm maps as observed by the 128-antenna configuration of the Murchison Widefield Array (MWA) Phase II. These visibilities are imaged to obtain ‘dirty images,’ which are then used to train a 2D convolutional neural network (CNN) to predict $\rm x_{HI}$. To systematically assess the effect of PSF mis-modelling, we generate multiple test sets by varying the MWA’s antenna layout, thereby introducing controlled variations in the PSF; we then feed these alternative PSF dirty images to our CNN trained using only dirty images with the PSF of the true antenna layout. Our results demonstrate that PSF variations introduce biases in the CNN’s predictions of $\rm x_{HI}$, with errors depending on the extent of PSF distortion. We quantify these biases and discuss their implications for the reliability of machine-learning-based parameter inference in 21-cm cosmology and how they can be utilized to improve the robustness of estimation against PSF-related systematics in future 21-cm surveys.  In concluding, we also discuss how this approach to incorporating realistic instrument error into an ML analysis pipeline can be expanded to include multiple other effects.}


\begin{document}
\maketitle
\flushbottom

\section{Introduction}
\label{sec:intro}
The redshifted 21-cm line of neutral hydrogen would serve as an excellent probe to understand the structure and evolution of our Universe across cosmic time. This signal provides a unique window into the key epochs in the evolution of our Universe, including the Dark Ages, Cosmic Dawn, and the Epoch of Reionization, offering deep insights into the formation of the first stars, galaxies, and large-scale structure.

However, observations of the 21-cm signal, both as a sky-averaged global signature and as fluctuations measured using an interferometer, are extremely challenging.  Detecting the cosmological 21-cm signal is one of the key science goals for several current and upcoming low-frequency radio telescopes (for example, SKA, \cite{Mellema_2013, Koopmans_2015}; HERA,\cite{DeBoer_2017_short}; MWA, \cite{Tingay_2013_short}; LOFAR, \cite{Harlem_2013_short}; etc.).
Several radio-interferometric experiments have reported upper limits on the power spectrum of 21-cm fluctuations over the past decade, from the epoch of reionization (EoR) \cite{patil_et_al_2017, barry_et_al_2019a, li_et_al_2019, trott_et_al_2020, Mertens_2020, hera_2021b, hera_2023, Mertens_2025} and the cosmic dawn (CD) era \cite{eastwood_et_al_2017, Gehlot_2019, Gehlot_2020, Yoshiura_2021, Munshi2023}. These upper limits have enabled theorists to rule out certain models of reionization and provide constraints on the physical parameters.

In recent years, machine learning (ML) techniques have become increasingly prominent across various different domains of astrophysics and cosmology. Particularly, in 21-cm cosmology, ML has been applied to a wide range of problems, such as signal emulation to parameter estimation, highlighting its potential as a powerful tool for analyzing complex high-dimensional datasets. For instance, fast emulators for the 21-cm global signal and power spectrum have been developed using ML approaches \cite{Cohen_2019, Schmit_2017}, enabling rapid exploration of parameter space. Convolutional Neural Networks (CNNs) have been employed to constrain reionization physics from 21-cm maps \cite{Hassan_2019, la_plante_and_ntampaka_2019}, while deep learning models have also been used to emulate the full time-evolution of 21-cm brightness temperature maps from the EoR \cite{Chardin_2019}. 
More recent advancements include the use of convolutional denoising autoencoders (CDAEs) to recover the EoR signal from simulated SKA images with realistic beam effects \cite{Li_2019}, and the application of Gaussian Process Regression (GPR) for foreground separation \cite{2024MNRAS.527.7835A}. Artificial Neural Networks (ANNs) have been used to estimate bubble size distributions from 21-cm power spectra \cite{Shimabukuro_2022}, and CNNs have been applied to 3D tomographic 21-cm images for parameter inference and posterior estimation \cite{Zhao_2022}. Also, \cite{Choudhury_2022} implemented ANNs to extract power spectrum parameters from synthetic observations contaminated with strong foregrounds. These diverse applications underscore the growing role of ML in tackling the multifaceted challenges of 21-cm cosmology.

In spite of the excellent works incorporating ML into EoR studies, ML methods have their drawbacks. All supervised-learning methods are highly problem-specific, strongly dependent on the quality, diversity and representativeness of the training process. They are mostly not generalizable and struggle to deal with out-of-distribution inputs or unknown systematics, common in real observations. While such models can perform well on narrowly defined tasks under controlled conditions, their reliability diminishes when faced with unmodelled complexities. ML becomes particularly useful when it becomes difficult to explicitly model the data, particularly where traditional physical models are  computationally expensive or analytically intractable. Its strength lies in its ability to learn from data representations directly, provided the training data sufficiently captures the variability of real observations.

To address these limitations, our work takes a step towards building a more resilient and informed ML model. In our case, to train a network, we need not define an explicit model for complicated systematics (a nearly impossible feat for instruments with the size and complexity of 21-cm cosmology experiments); we just need a way of representing a reasonable range of signatures in our training set. As a first case study, we explore the impact of PSF variability on ML-based inference, highlighting the need for more robust and systematics-aware training strategies in 21-cm cosmology.

In real data, all the complicating factors are present.
To confidently analyze real data with ML, we need an approach that will not confuse one issue for another.  For instance, miscalibrated foreground removal and electromagnetic mutual coupling between antennas can both leave excess power at the same spectral scales (due to their comparable light travel time delays). An ML approach trained only to remove foregrounds is unlikely to do the job properly if mutual coupling is present.  There is no existing complete end-to-end pipeline stitching all of these key aspects together. The work presented here is a first step in a series of papers, through which we  will methodically incorporate more and more sophisticated effects into an ML-based end-to-end pipeline to analyze 21-cm interferometric data. In this paper, we aim to demonstrate how well we can predict the neutral hydrogen fraction directly from dirty images, bypassing the necessity to reconstruct the original image after going through a deconvolution algorithm.

The outline of this paper is as follows: In \textsection\ref{sec:21cm signal}, we discuss briefly the fundamentals of interferometric measurements, focusing on visibilities and the role of the point-spread function (PSF). In \textsection\ref{sec:methodology}, we describe our approach to generate the mock-observations for preparing the training set, and how simulated 21-cm maps and visibilities are generated. We also describe the architecture of the neural network model that is used. Following this, we describe how the PSF errors are introduced and the test sets are prepared. In \textsection\ref{sec:results} we present our results and in \textsection\ref{sec:discussions and conclusions} we summarize and discuss the findings of this paper. Finally in \textsection\ref{Roadmap for the future}, we further chalk out the roadmap for future work in detail.

\section{The 21-cm signal and interferometric measurements}
\label{sec:21cm signal}
The key observable of 21-cm cosmology is the differential brightness temperature $\delta T_b$, which primarily quantifies the contrast between the redshifted 21-cm signal from neutral hydrogen and the cosmic microwave background (CMB). The differential brightness temperature of the 21-cm signal $(\delta T_b)$ at a redshift $z$ and an angular position $x$ can be expressed as:
\begin{equation}
    \delta T_b(x,z) \approx 27 x_{\text{HI}}(x,z) [1+\delta
(x,z)] \left(\frac{\Omega_b h^2}{0.023}\right)\left(\frac{0.15}{\Omega_mh^2}\frac{1+z}{10}\right)^{1/2} \left(\frac{T_s(x,z) - T_{\text{CMB}}(z)}{T_s(x,z)} \right) \text{mK},
\label{eqn:deltaTb_full}
\end{equation}
where, $x_{\text{HI}}$ is the neutral hydrogen fraction, $\delta$ is the matter overdensity contrast and $T_s$ is the spin temperature, respectively of the region located at $(x,z)$. $T_{\text{CMB}}$ is the cosmic microwave background (CMB) temperature which is the radio-background temperature at redshift $z$.  $\Omega_m,\Omega_b$ are the mass and baryon densities in units of the critical density respectively. For a more detailed review see \cite{Zaroubi_2013, Furlanetto_2006}. Assuming very efficient x-ray heating at a high spin temperature limit, with $T_s>>T_{CMB}$, Eq.\ref{eqn:deltaTb_full} can be simplified as:
\begin{equation}
    \delta T_b (x,z) \approx x_{\text{HI}}(x,z) [1+\delta(x,z)]\left(\frac{1+z}{10}\right)^{1/2}\text{mK}.
\end{equation}

Eq.\ref{eqn:deltaTb_full} highlights the sensitivity of the 21-cm signal to a wide range of astrophysical and cosmological parameters. Throughout this paper, we have used the Planck 2018 cosmology \cite{planck_2018_VI}.  
However, the radio-interferometers do not observe $\delta T_b$ directly. Instead, they measure complex-valued quantities called visibilities, which represent baseline-dependent, spatial Fourier modes of the sky, modulated by the instrument's primary beam response.

Assuming that the sky is effectively flat in the region of interest (we plan to deal with the curved sky in future work), the measured visibility function $ V(u,v) $, is related to the sky brightness distribution $ I(l,m) $ via a 2D Fourier transform:

\begin{equation}
V(u,v) = \int A(l,m) I(l,m) e^{-2\pi i (ul + vm)} \, dl \, dm
\end{equation}
where, $ I(l,m)$  is the sky brightness distribution and $A(l,m)$ is the primary beam pattern of the antenna as a function of angular coordinates $( l, m )$, $( u,v )$ are the spatial frequency coordinates in units of wavelengths, and the exponential term represents the phase difference due to different arrival times of the signal at each antenna. Longer baselines (larger  $( u,v )$ values) probe smaller angular scales, while shorter baselines capture large-scale structures. 

Observations of the 21-cm signal would provide us with visibilities, which can be imaged to reconstruct spatial maps of the redshifted 21-cm signal, providing a more informative view of the distribution of neutral hydrogen across redshifts. While the power spectrum is the primary statistical observable targeted by current interferometric experiments, direct imaging would offer far richer insights into these high redshift epochs. Direct imaging would be a rich, treasure trove of information, and would enable us to resolve individual ionized and neutral regions, revealing the morphology of reionization and the large-scale structure of the early universe.

However, due to the faintness of the 21-cm signal and contamination from foregrounds like Galactic synchrotron emission, direct imaging is significantly more challenging. While statistical detection via the power spectrum from the current and upcoming interferometers is expected first, direct imaging will provide a transformative leap in our understanding of the cosmic dawn and reionization epochs.
The SKA is predicted to achieve the sensitivity required for the direct imaging of the signal and eventually enable us to map the tomography of the signal. 

\subsection{The point spread function}
A fundamental concept in interferometric imaging is the point spread function (PSF), which characterizes how a point source appears in the reconstructed sky image due to the instrument's limited sampling of spatial frequencies. It is the response pattern corresponding to the instrument's size and sampling. Since an interferometer does not measure all possible Fourier modes, the resulting image, which is called the `dirty image,' is a convolution of the true sky brightness with the PSF (which is also known as the `dirty beam').

The PSF is determined by the \textit{uv-coverage}, which describes the sampled spatial frequencies based on the distribution of the antennas and the observation time. A well-sampled uv-plane results in a more compact and symmetric PSF, leading to higher-quality imaging, while sparse coverage introduces sidelobes, artifacts that can complicate image interpretation. 
This convolution of the sky with the PSF, introduces spatially varying distortions that can mimic or obscure cosmological features in 21-cm maps.  Consequently, understanding and mitigating PSF effects are critical for robust inference. Also, understanding the structure and limitations of the PSF is essential when interpreting interferometric images or using them as inputs for inference pipelines. It is to be kept in mind that CLEAN or any deconvolution technique is non-linear, and can lead to loss of information. Hence, working directly with the `dirty image', which is the truest representation of the measurements, becomes more beneficial. 

In real observations, the PSF is not fixed; it varies with time, frequency, and array configuration. Accounting for such variations into inference pipelines is essential because models trained on a single, idealized PSF might not perform well, when applied to slightly different conditions.  In this paper, we investigate this effect in a controlled setting.  As the first step, we test to verify that the mere presence of a fixed PSF does not adversely affect the recovery of the neutral hydrogen fraction $(\rm x_{HI})$ directly from the dirty images.  Only after establishing this baseline do we proceed to study how variations in the PSF influence the inference. We then try to mimic variations introduced in the PSF corresponding to observations from the 128 antenna configuration of the MWA Phase II array by systematically removing randomly selected antennas from the default configuration. Each of these cases give rise to a slightly different measurement of the same field being observed, and a slightly different PSF. Our goal is to quantify how much variation in the PSF can be tolerated before the accuracy of parameter inference from these mock observations begins to degrade significantly. This analysis provides a first step toward understanding the level of error that needs to be represented in the training sets to construct an ML-based pipeline that is robust to realistic instrumental systematics.

\section{Methodology}
\label{sec:methodology}
To develop our ML-framework, the first and most crucial step is assembling a representative and well-characterized training set. In our case, this involves generating 21-cm maps and obtain their corresponding dirty images created using a reference telescope configuration. The ML-framework is then trained on these ideal dirty images, where the PSF is consistent and well-behaved across all training samples. This allows the network to learn the mapping between the observed image and the underlying neutral fraction. However, in real observations, small changes in the telescope array—such as antenna failures or flagged baselines—can alter the PSF in ways that deviate from the idealized case. To test the resilience of the trained model under such realistic conditions, we introduce controlled perturbations to the PSF in the test set by randomly removing antennas from the default configuration. This setup allows us to systematically quantify how much PSF variation the model can tolerate before its predictive performance begins to degrade.  We explain our training data generation and model architecture in detail in this section.

\subsection{The 21-cm Maps}

We first assemble a suite of simulated 21-cm signals to build our training sets.
To date, the theoretical interpretations of the 21-cm measurements have mostly been guided by semi-numerical models of reionization \citep{Mesinger_2011, Santos_2010, Fialkov_2014, hutter2018}.  Other fast approximate techniques have been developed \citep{Thomas_2009, Ghara_2015} to bypass the more computationally expensive and accurate full radiative transfer simulations \citep{gnedin2014, rosdahl2018, Ocvirk2020, Kannan2022}.
While these simulations provide us an excellent representation of various possible reionization scenarios, they are all subject to their own set of limitations and approximations which make it difficult to incorporate all of them simultaneously into an inference framework.

Our initial result presented here uses only simulations from \texttt{21cmFAST} v3 \citep{murray_et_al_2020}. We have generated 1000 lightcones, each with different reionization histories. A lightcone is a spatial and temporally coherent 3D volume, encoding the evolving 21-cm brightness temperature field as a function of redshift, along the line of sight.  Each lightcone consists of a sequence of 2D slices corresponding to discrete redshifts, with each slice encoding the spatial distribution of the 21-cm brightness temperature and the associated neutral hydrogen fraction at that redshift. 
To construct our training set, we randomly sample 2D slices from these 1000 lightcones along with their neutral hydrogen fraction at a single frequency. 

These 21-cm maps and the corresponding $x_{\rm HI}$ values serve as the foundation of our supervised learning pipeline. We emphasize that, in this work, we limit ourselves to individual 2D slices at fixed frequencies, deferring a full frequency-dependent analysis to a forthcoming paper in this series.

\begin{figure}[t!]
    \centering
    \includegraphics[width=\linewidth,trim=0cm 1cm 0cm 2cm, clip=True]{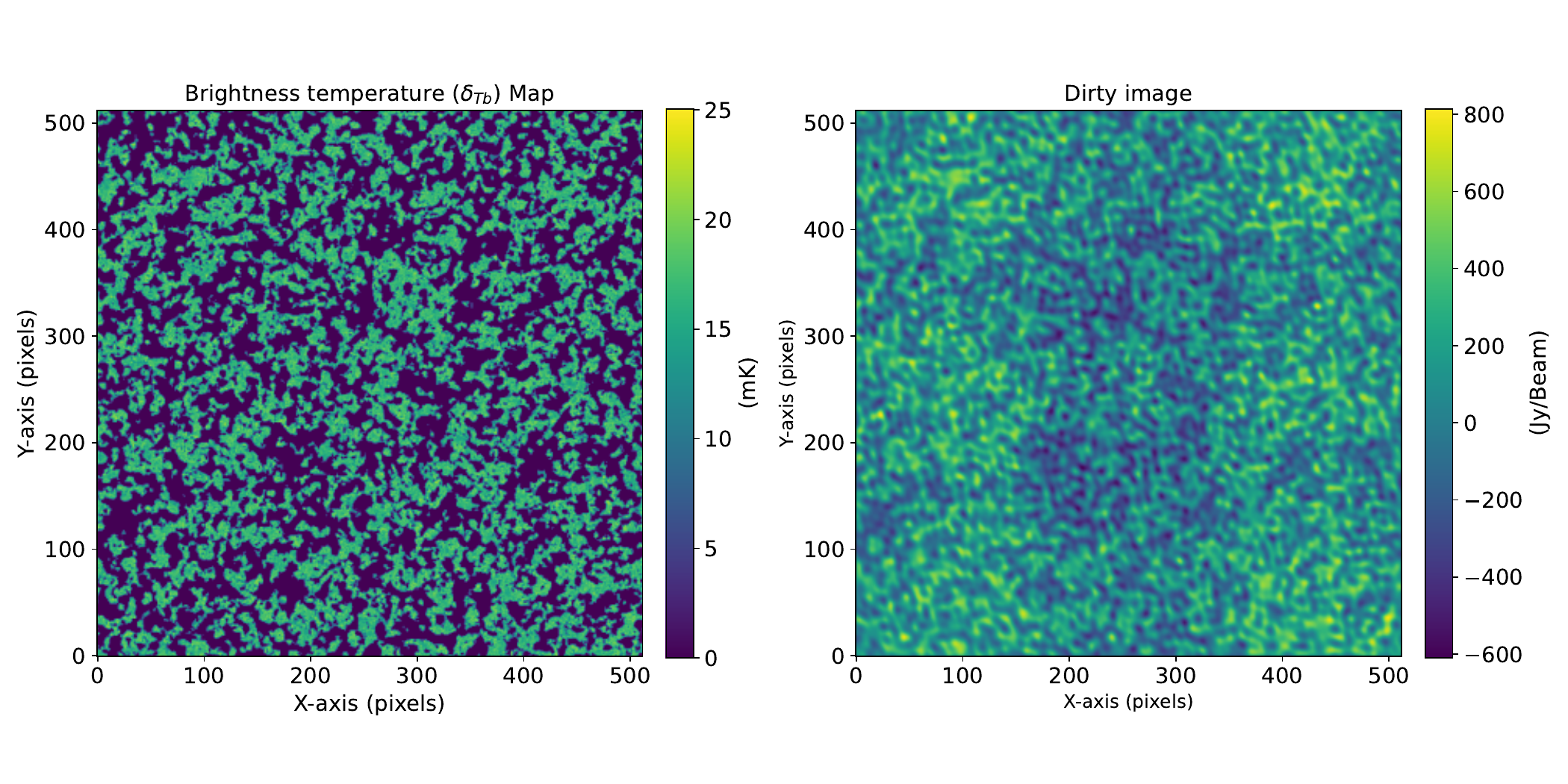}
    \caption{\emph{Left}: One slice of a 21-cm brightness temperature field from a \texttt{21cmFAST} simulation.  \emph{Right} A ``dirty image'' of same slice, made by creating model visibilities from the left hand image (via CASA's \textit{simobserve}) and then imaging those visibilities (via CASA's \textit{tclean}).}
    \label{fig:21-cm-images}
\end{figure}
\subsubsection{Simulating visibilities \& Imaging}
Once we have the set of randomly chosen slices from different lightcones,
the next step of our analysis is to put the 21-cm signal maps through an instrument model, produce interferometric visibilities, and then image those visibilities for input into our ML network.  We choose this approach (as opposed to simply convolving the images with a PSF calculated from the uv-coverage) as a foundation for future work, where other visibility-based systematics will be considered.

The MWA Phase II \citep{wayth_et_al_2018} is an upgraded version of the original Phase I array. We consider the MWA Phase II compact configuration as our model instrument, with a tightly packed core of radius 50 m, and additionally, the two hexagons for calibration and power spectrum sensitivity. The majority of the baselines in the Phase II configuration are less than $200$\,m, and the redundancy present in the hexagons leads to grating lobes in the PSF.  We choose to work with this configuration as something as a worst-case scenario for the PSF present in a tomographic imaging experiment.  We use the “EoR0” field (Right Ascension, (R.A.)=0h00, Declination (decl.)=-27°) for 10800s (three hours) and use CASA's \textit{simobserve} task to create single-frequency model visibilities for the brightness temperature $(\delta T_b)$ image slices pulled from each of the light cones.  
We then, in turn, use the \textit{tclean} task, with the number of iterations set to zero (i.e. there is no CLEAN process), to transform these visibilities to dirty images of the 21-cm signal.  An example of this process is shown in Figure \ref{fig:21-cm-images}.

There are sophisticated codes which particularly model the curved sky accurately, which we plan to deal with in our future papers. For example, codes such as \texttt{pyuvsim} \citep{lanman_et_al_2020} can be used for generating visibilities and FHD \citep{sullivan_et_al_2012,barry_et_al_2019a} for map-making. 

\subsection{ML Network}
\label{sec:cnn}

The training set for our 2D Convolutional Neural Network (CNN) contains approximately $5000$ dirty images of the 21-cm signal. These slices have been randomly picked up from the suite of 1000 lightcones that we had generated. We consider only those slices with neutral hydrogen fractions in the range $(0.10, 0.90)$, so as to avoid blank or featureless images. Each dirty image is associated with a corresponding neutral hydrogen fraction, $x_{\rm HI}$, which represents the abundance of neutral hydrogen in that specific slice before passing the image through the instrument model. The CNN is trained to predict the neutral hydrogen fraction $x_{\rm HI}$ directly from these dirty images.

Our CNN model is designed with two convolutional layers, each followed by batch normalization and max pooling layers. These convolutional layers allow the network to capture spatial patterns and hierarchical features in the images, while batch normalization helps in stabilizing and accelerating the training process. Max pooling layers are employed to progressively reduce the spatial dimensions and make the network invariant to small shifts in the input, enhancing its ability to generalize. At the end of the convolutional stack, the output is flattened, and a fully connected (dense) layer serves as the output layer, which aggregates the learned features to predict  $x_{\rm HI}$ for each image. 
\begin{figure}[t!]
    \centering
    \includegraphics[width=\linewidth,trim=0cm 1cm 0cm 2cm, clip=True]{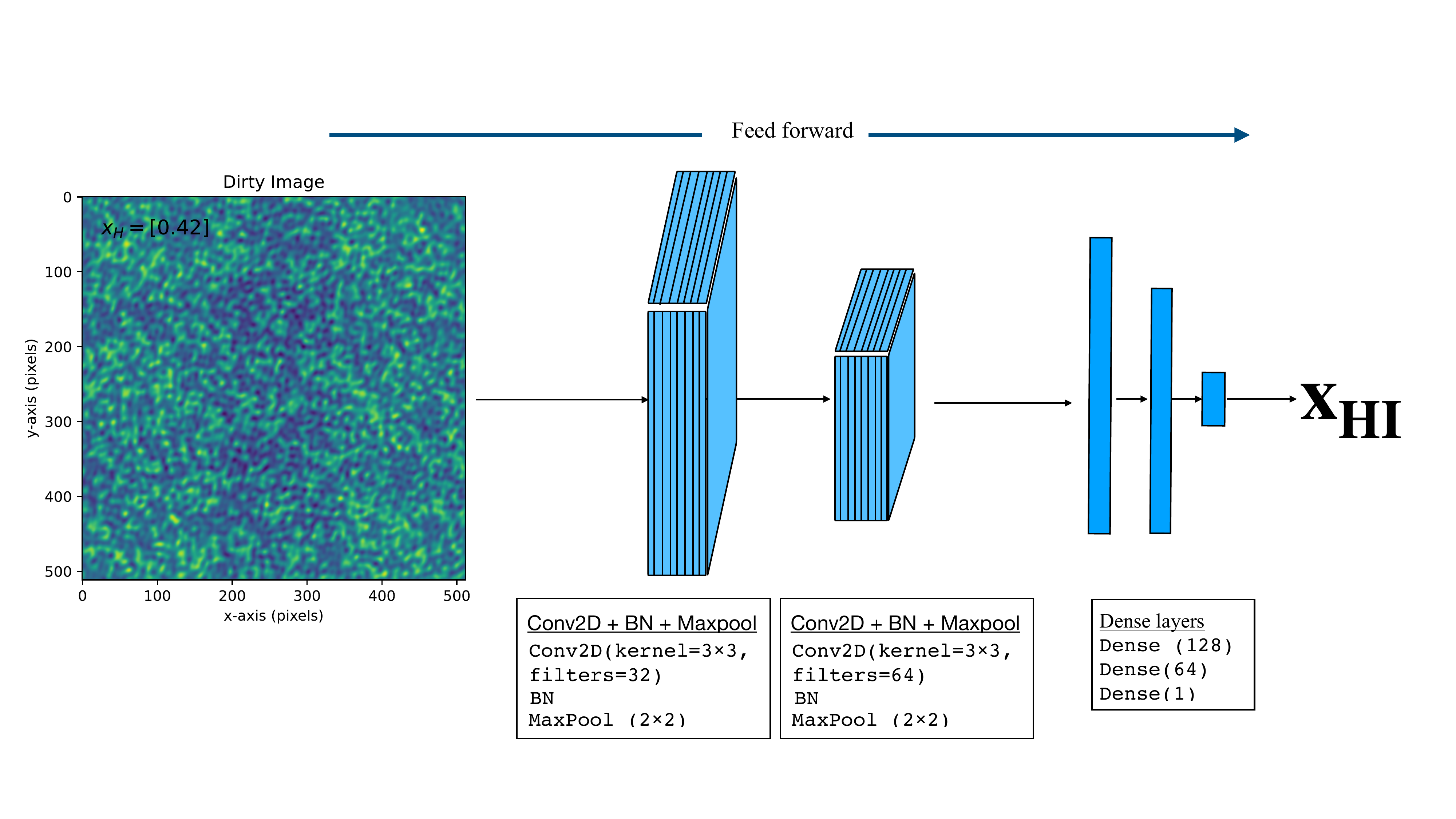}
    \caption{ A schematic representation of the 2D CNN architecture. The input images are the `dirty images' generated using CASA's \textit{simobserve} and \textit{tclean} tasks. The CNN consists of two sets of convolutional layers, batch normalization and maxpool layers which finally regresses onto the neutral hydrogen fraction $x_{\rm HI}.$}
    \label{fig:CNNarchitecture}
\end{figure}

Our CNN model (Fig.\ref{fig:CNNarchitecture}) is trained for 500 epochs using a mean squared error loss function, which measures the difference between predicted and true values of 
$x_{\rm HI}$. The training process optimizes the model weights to minimize this error(we use the Adam \citep{Kingma_2014} optimizer from \textsc{keras}), gradually refining the network’s ability to accurately estimate the hydrogen fraction from the corrupt 21-cm image data. We split the training data into two subsets, allocating $20\%$ of it as the validation set. This is done to monitor the model’s performance, prevent overfitting and ensure generalization. We have used publicly available python-based packages \textsc{scikit learn} (\cite{Pedregosa_2011}) and \textsc{keras} in this work to design the networks. 
We train the CNN-model to learn spatial patterns within the images that correlate with variations in $x_{\rm HI}$, effectively capturing the underlying structure and distribution of the 21-cm signal in a mock-observational environment. This model effectively learns the underlying relationship of the observed images and  physical parameters, enabling more accurate interpretations of 21-cm signal data in future cosmological studies.

\section{Results}
\label{sec:results}
In this section, we present the results of training the CNN-model on the $\delta T_b$ maps, convolved with the default PSF of the full MWA Phase II array, as described in \ref{sec:cnn}. We train the CNN to learn the non-linear mapping between these `dirty' images and the corresponding neutral hydrogen fraction ($\mathrm{x_{HI}}$).

We create a test set consisting of 200 dirty images slices, following the same steps as the training data, along with their corresponding $\rm x_{HI}$. This set is different from the training data, and remains unseen to the trained network.
For evaluation, we compare the CNN-predicted values of $x_{\rm HI}$ against the true values from the test set using two standard metrics: the coefficient of determination ($R^2$) and the root mean square error (RMSE). The $\rm R^2$ score and RMSE are defined as:
\begin{equation}
    \mathrm{R^2=1-\frac{\Sigma(x_{HI,pred}-x_{HI,ori})^2}{\Sigma(x_{HI,ori}-\overline{x}_{HI,ori})^2}}
\end{equation}
\begin{equation}
    \rm RMSE = \sqrt{\frac{1}{N}\sum_{i=1}^N \left(\frac{x_{HI, ori}-x_{HI, pred}}{x_{HI, ori}}\right)^2}
\end{equation}
where, $\mathrm{{x}_{HI,ori}}$ is the original parameter, $\mathrm{x_{HI,pred}}$ is the parameter predicted by the CNN, $\mathrm{\overline{x}_{HI,ori}}$ is the average of the original parameter, and the summation is over the entire test set, consisting of N samples. $\mathrm R^2$ can vary between 0 and 1, and $\mathrm R^2 =1$ implies a perfect inference of the parameters. A low value of RMSE implies a good prediction.

\begin{figure}[t!]
    \centering
    \includegraphics[width=0.99\textwidth]{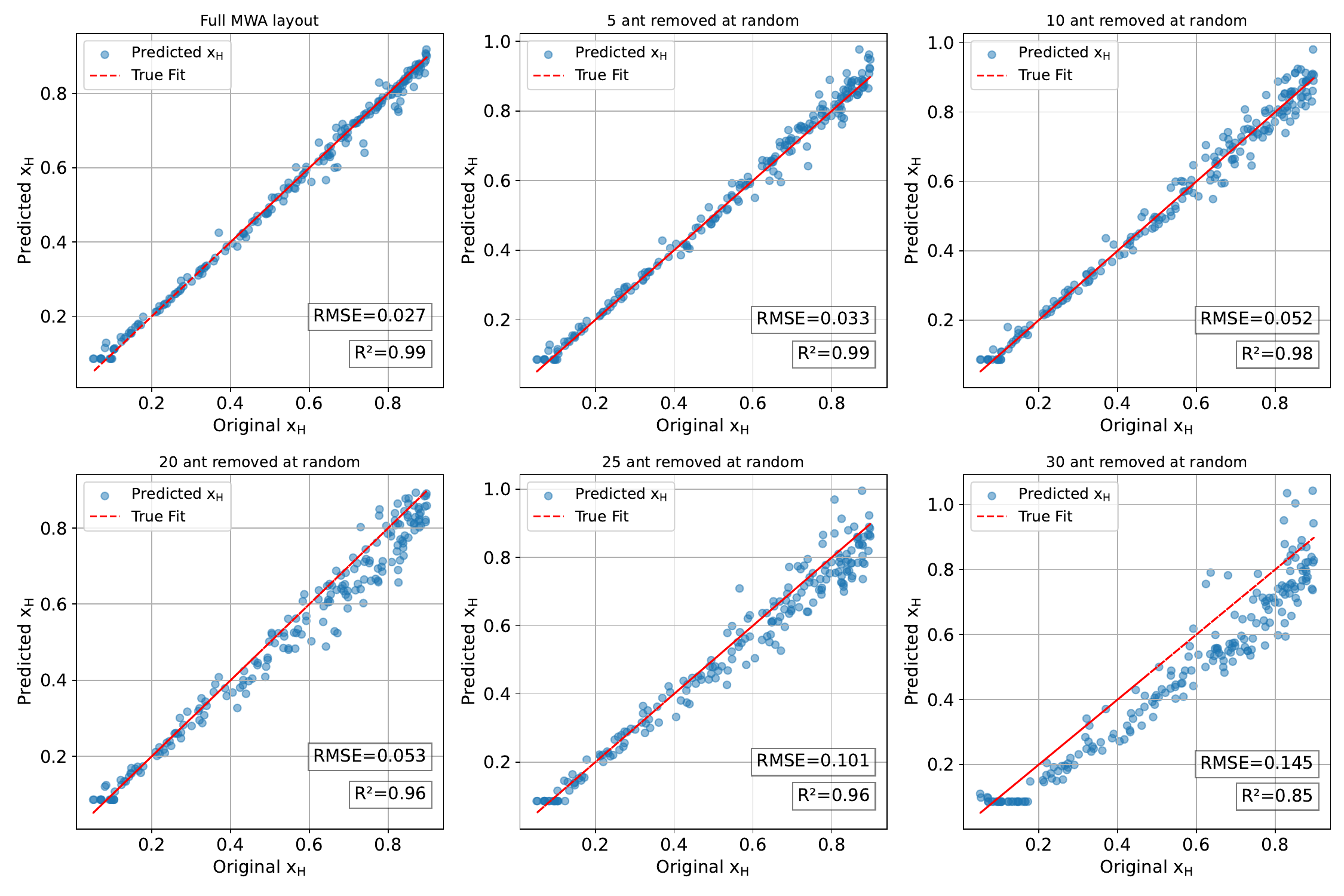}
    \caption{The predictions of the neutral fraction ($x_{\rm HI}$)from our CNN with different testsets. The first plot on the left panel corresponds to the results obtained from a testset with no PSF variations introduced. }
    \label{fig:scatterplots}
\end{figure}

The top-left plot in Fig. \ref{fig:scatterplots} shows the results from CNN. We plot the true versus the predicted values of $x_{HI}$ along with the computed $\rm R^2$-score, for a test set with no variations in the PSF introduced. The CNN achieves an excellent $\rm R^2$ score and low RMSE, demonstrating that the neutral fraction can be robustly inferred from dirty images when the PSF is fixed and well-characterized. This confirms that the convolution with a known PSF does not significantly degrade the information content needed for learning $x_{\rm HI}$.

We would like to investigate whether a CNN trained on a certain PSF (i.e. the PSF of the full MWA Phase II with all 128 antennas) could also correctly recover the neutral fraction of these maps where the PSF is slightly different. The scope of this test is limited to differences within a reasonable range: we are not attempting to generalize across entirely different instruments (e.g., training on MWA and predicting for SKA) nor to train on PSF-free images and then recover parameters from realistic data. Instead, our goal is to establish whether a CNN trained on a well-defined PSF can remain robust when faced with modest, realistic perturbations that reflect the level of uncertainty expected in an observing pipeline where instrumental and observational parameters are known to decent accuracy.
Having established this baseline performance, we next test the robustness of the CNN predictions under degraded PSFs that mimic calibration uncertainties and array-configuration variability.

\subsection{Introducing PSF Errors}
In real interferometric observations, the PSF is intrinsically frequency dependent and evolves with the UV-coverage, which itself varies with observation time, antenna configuration, flagging, calibration imperfections and several other factors. These corruptions creep into the image domain as artifacts at that can in principle mimic or obscure the features in the cosmological signal. When we train an ML pipeline on dirty images assuming a perfect PSF, the pipeline might not be capable of dealing with real-life interferometric observations. Hence, we introduce imperfections in the assumption of the PSF of the observing instrument by removing random subsets of antennae from the ideal MWA Phase II configuration. In this way, we can mimic one aspect of instrument model uncertainty in a very simplistic manner.

In this preliminary analysis, we focus on uncertainty in the PSF.  To introduce errors into the PSF, we rerun \textit{simobserve} on a new set of \textbf{$\delta T_b$} maps, but randomly remove antennae from the array to degrade the PSF. Randomly removing antennae would modify the \textit{uv}-sampling and hence introduce variations in the PSF, simulating a mock-observation scenario. We take the original $\delta T_b$ slices corresponding to the test set, and create another set of visibilities after randomly removing 5 antennas from the 128 antennas composing MWA Phase II.  Each slice within a set has a different set of $5$ randomly chosen antennas removed from the default configuration, so that the PSF error is different in each sample.  
We then repeat this process, by removing 10, 20, 30, and 64 antennas, for a total of five such test sets with 200 visibility simulations per set using the \textit{simobserve} task from CASA.  We then re-image these visibilities with the \textit{tclean} task. 
These sets of `dirty images' for different sets of antennas removed from the default MWA phase II configuration now constitute different test sets, with slightly modified instrument configurations. We use our previously trained CNN pipeline to predict the neutral hydrogen fractions $\rm (x_{HI})$ from these test sets which are representative of dirty images created by modified PSFs. In Fig. \ref{fig:scatterplots}, we plot the results obtained from each of the test sets described above. The scatter plots show the true versus the predicted neutral hydrogen fraction, where each point is the predicted $\rm x_{HI}$ from one of the 200 dirty image samples in the test sets, and corresponds to a different overall antenna layout.

\begin{figure}[t!]
    \centering
    \includegraphics[width=0.85\textwidth]{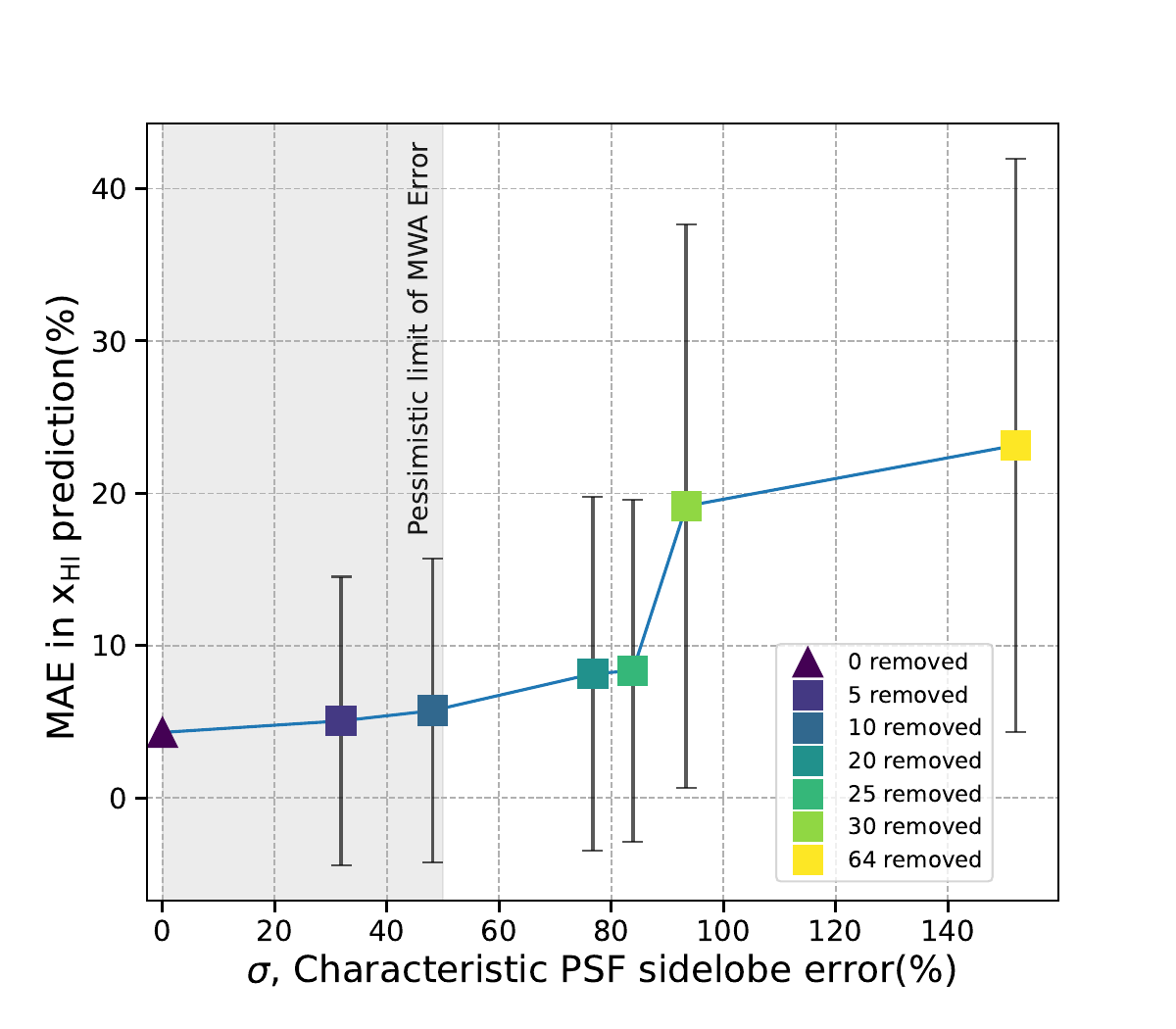}
    \caption{The percentage error in our CNN's predictions of the neutral fraction ($x_{\rm HI}$) as a function of the characteristic PSF error compared with the full 128-antenna PSF that was used in the training set.  The shaded grey region indicates the most pessimistic range of PSF error we expect with the real MWA, suggesting that a training set containing only the full 128-antenna PSF is sufficient to yield $\sim5\%$ error in neutral fraction predictions even we are mismodeling the instrument's true PSF. The error bars here denote the standard deviations of the absolute fractional errors on the $\rm x_{HI}$ predictions for all the samples, in each of the test sets. }
    \label{fig:metrics from psf-varied test sets}
\end{figure}

From Fig.\ref{fig:scatterplots}, it is clear that the predictions for the same underlying $\delta T_b$ maps and their associated $\rm x_{HI}$ become worse as more variations are introduced in the PSF, by removing more number of antennae. To capture and quantify the error introduced into the PSF, arising from variations in the default array configuration, we define a metric based on the root mean squared error (RMSE) between the perturbed PSF and a reference `true' PSF (computed from the default, MWA Phase II full array configuration). Specifically, for each realization, the observed PSF is saved after randomly dropping a subset of antennas, and the pixel-wise RMSE is computed relative to the true (default) PSF. This RMSE error is then normalized by the total pixel-wise intensity of the true PSF to obtain a dimensionless score. \\
The RMSE based error metric, $\epsilon_{\rm PSF}$, for each test image, is defined as:

 \begin{equation}
     \rm \epsilon_{PSF}= \sqrt {\frac{1}{N} \sum _{i=1}^N\left( \frac{B^{obs}_i -B^{ori}_i}{B^{ori}_i} \right) ^2}
 \end{equation}

where, $\rm B_i$ is the PSF response at pixel $i$ and $N$ denotes the total number of pixels in each image. This score enables us to quantify variations between the true and modified or perturbed PSFs for each image in the test set.
 
The final score of the PSF error metric $\sigma$ for each of our test set is calculated as follows. Our error metric $\rm \epsilon_{PSF}$, computes the normalized difference squared between the original PSF and the modified PSF, which is then averaged over all the pixels, N, in the image. This is then averaged across all the 200 modified PSF's corresponding to the test sets for that specific number of antennas removed. This quantity, $\sigma$, encapsulates the characteristic PSF sidelobe error for a particular test set and is given by:

 \begin{equation}
     \rm \sigma=\frac{1}{M}\sum_j^{M} \epsilon_{PSF}^j
 \end{equation}
where, M is simply the number of test set samples (200) for each case considered (5,10,20,30,64 antennae removed) and we obtain the $\sigma$ corresponding to each test set.

Now that we have a metric to represent the characteristic sidelobe errors for each test set, we also compute the mean absolute error (MAE) for the predictions of the CNN corresponding to each of these test sets. The MAE for the predicted $\rm x_{HI}$, for each of the test sets is given by:

\begin{equation}
     \rm MAE = \frac{1}{M}\sum_{i=1}^M\frac{|x_{HI,i}^{pred}-x_{HI,i}^{ori}|}{x_{HI,i}^{ori}},
\end{equation}

This number represents the typical error in the prediction of $x_{\rm HI}$, for each of the test sets with $0,5,10,20,30,64$ antennae removed. In Fig.\ref{fig:metrics from psf-varied test sets}, we see that this error correlates strongly with the computed PSF error metric, for each of the test cases considered. The shaded grey region shows the worst case PSF error we might expect in real MWA observation, calculated based on the typical number of malfunctioning antennas in an MWA observation \citep{joseph_et_al_2020}.  This analysis suggests, then, that using only the full 128-antenna PSF in the training is sufficient to yield $\sim5\%$ uncertainties in $x_{\rm HI}$ even if the true PSF differs because of real-world effects in the instrument. We also note that our CNN typically only recovers $x_{\rm HI}$ with $\sim4\%$ error even when the PSF is modelled perfectly, suggesting that the model itself is the limiting factor at these low levels of PSF error. In addition, we also calculate the standard deviation of the normalized residuals,   $\xi=(\rm x_{HI,ori}-x_{HI,pred})/x_{HI,ori}$ corresponding to each test set containing 200 samples, and plot them as error bars [\ref{fig:metrics from psf-varied test sets}]. The standard deviation of the normalized residuals are calculated as, $\rm \sqrt{\frac {1}{M}\sum{(\xi-\overline{\xi})^2}}$. These give us a measure of the possible deviation from the true values of the parameters within the same test set, demonstrating how the positions of the antennae removed also is reflected in the variations of the PSF and are plotted as the error bars in Fig.\ref{fig:metrics from psf-varied test sets}.

\section{Discussions}
\label{sec:discussions and conclusions}
Our results demonstrate that, within the range of PSF variations considered here, a CNN trained on dirty images convolved with the default PSF can still recover the global neutral hydrogen fraction, $\rm x_{HI}$, with high accuracy. While the presence of PSF deviations does introduce a measurable increase in prediction error, these deviations which arise from randomly removing subsets of antennas in the array, remain within acceptable limits.  Overall, this indicates that for realistic, modest PSF errors, training exclusively on unperturbed PSFs is sufficient for reliable inference of 
$\rm x_{HI}$.
However, the variation in the MAE \% error reflects not only the sensitivity of the predictions to the overall strength of the PSF perturbations, but also the fact that different random subsets of removed antennas produce distinct PSFs. This randomness in array configuration leads to a spread in prediction performance, which is captured in the error bars.

This highlights the sensitivity of ML-based inference pipelines to PSF-induced distortions, even when the PSF perturbations are relatively simple and physically plausible. The approach presented here can be viewed as a first controlled step toward this broader goal, serving as a testbed for developing error-aware and uncertainty-aware inference models. 
This is a very simple, proof-of-concept demonstration of the ability of an ML based pipeline to directly estimate physics parameters from mock-observations. The variations introduced in this work are deliberately simplistic: random antenna removals simulate a subset of real-world effects such as antenna flagging or hardware failure. These assumptions are sufficient to demonstrate the susceptibility to error, of ML models trained under idealized assumptions, underscoring the need to incorporate more realistic instrumental systematics during training. We do not consider other complex factors that might introduce errors in the modelling of the PSF, and thereby introduce artefacts in the observed images.

\section{Roadmap for the future}
\label{Roadmap for the future}
Most of the inference pipelines in EoR experiments do not explicitly account for the errors introduced by time and frequency-dependent instrumental characteristics, which distort real interferometric observations. In this work, we have considered the PSF at a single frequency only, 
focusing on 2D image slices, which represent only a simplified subset of the information available in 21-cm observations. While this has provided a useful first step for exploring the effect of PSF variability on CNN-based inference, a critical next stage is to extend the framework to include the full frequency (or redshift) axis. Doing so will allow the models to capture the rich temporal and spectral evolution of the 21-cm signal, which is indispensable for extracting astrophysical parameters in practice from interferometric observations.

Equally important will be the systematic incorporation of noise and calibration uncertainties. Thermal noise, ionospheric distortions, and residual calibration errors all imprint characteristic signatures in the data that could strongly influence inference pipelines. A natural progression from this work would therefore involve training and testing models in the presence of these additional systematics, moving incrementally toward a realistic observational regime.
A further challenge is the treatment of foregrounds, which are many orders of magnitude brighter than the cosmological 21-cm signal. Integrating foreground into the ML framework, either as contaminants to be strategically modelled and subtracted or as part of a joint inference strategy, will be an essential milestone.
Finally, bridging the gap between idealized training and real data will require moving from flat-sky image slices to curved-sky representations, such as HEALPix maps. Incorporating these into end-to-end pipelines would enable the community to test ML inference under conditions that more closely approximate the complexities of actual experiments like the MWA and, eventually, the SKA. Stitching together all these components---frequency dependence, noise and calibration effects, foregrounds, and curved-sky geometries---will lay the foundation for robust, error-aware machine-learning pipelines capable of delivering reliable astrophysical constraints from future 21-cm surveys.
In the series of papers that will follow, our primary goal will be to develop ML pipelines that can ingest realistic dirty images---including direction-dependent effects, beam chromaticity, calibration errors, and other systematics---and are able to produce reliable astrophysical predictions.

\acknowledgments

MC is supported by a CFPU Postdoctoral fellowship.  Part of this research was conducted using computational resources and services at the Center for Computation and Visualization, Brown University. 


\bibliographystyle{JHEP}
\bibliography{totalref}

\end{document}